\newcounter{floating}
\def\thefloating{\@arabic\c@floating}
\def\fps@floating{tbp}
\def\ftype@floating{1}
\def\ext@floating{lof}
\def\fnum@floating{~}
\def\floating{\@float{floating}}
\let\endfloating\end@float
\@namedef{floating*}{\@dblfloat{floating}}
\@namedef{endfloating*}{\end@dblfloat}
\documentstyle[twocolumn,float,revtex]{aps}

\begin{document}
\begin{narrowtext}\widetext

\twocolumn
\begin{floating}[t] \widetext
\draft
\preprint{DRAFT: \today}
\begin{title}
Spin-Peierls ground states and frustration 
in a multi-band Peierls-Hubbard model
\end{title}

\author{H. R\"oder$^{(a)}$, A.R.~Bishop$^{(b)}$, and J.~Tinka Gammel$^{(c)}$}
\begin{instit}
$^{(a)}$ Physikalishes Instit\"ut, Universit\"at Bayreuth,
W-8580 Bayreuth, F.R.G. \\
$^{(b)}$ Theoretical Division,
Los Alamos National Laboratory, Los Alamos, New Mexico 87545 \\
$^{(c)}$ Materials Research Branch, NCCOSC RDT\&E Division (NRaD), 
San Diego, CA 92152-5000
\end{instit}

\begin{abstract}
We discuss the consequences of including both electron-phonon
and electron-electron couplings in multi-band models, focusing
on numerical studies of a one-dimensional two-band model in
the intermediate regime for both coupling strengths.
Spin-Peierls as well as  long-period, frustrated
ground states are identified, reminisent of those found in
antiferromagnetic next-nearest neighbor (ANNNI) models.
We speculate on experimentally observable signatures
of this rich phase diagram.
\end{abstract}
\end{floating}

\narrowtext

Multi-band and multi-orbital models have recently been much
studied, primarily due to their relevance for high-temperature
superconductors (HTC) \cite{gen_HTC}. At stoichiometry,
such models can exhibit interesting and unusual ground
states, such as electron-phonon ($e$-ph) driven incommensurate
long-period (LP) or superlattice (SL) phases \cite{batistic_long_per_91}.
Here we show that a one-dimensional (1D), two-band (2B) model
with competing electron-electron ($e$-$e$) and $e$-ph
interactions also exhibits at stoichiometry
interesting {\it magnetic} behavior: namely, LP frustrated or
spin-Peierls (SP) phases in the intermediate
regime between the strongly electron-electron ($e$-$e$) correlated
antiferromagnetic (AF) limit and the strongly $e$-ph correlated
large lattice distortion (LD) limit. Similar complex LP phases are
found in antiferromagnetic next-nearest neighbor (ANNNI) models
with competing magnetic interactions \cite{annni}.
Doping into such SP phases may induce unusual pairing mechanisms
in 1D or in the 2D version of this model,
relevant to HTC \cite{zhang_rice,emery,yonemitsu,imada}.

Considerable effort has gone into solving effective
one-band (1B) models. For dominant $e$-$e$ interactions
Zhang and Rice \cite{zhang_rice} derived in the context of HTC,
due to the separation of energy scales, an effective 1B $t$-$J$
Hamiltonian using a Wannier singlet basis. Imada \cite{imada} has
suggested that inclusion of an $e$-ph dependence of the effective
spin interaction $J$ is crucial, as it leads to the opening of a
spin gap, important for singlet superconductivity.
Here, we stress phenonena for which keeping the {\it full} 2B model
with {\it both} $e$-$e$ and $e$-ph interactions is essential.
These effects are generic in the
sense that their existence is not dependent on the exact analytic
form of the couplings or the specific parameters used.

\begin{floating}[t] \vskip 7.5truecm\end{floating}

We study the 1D, 2B, 3/4-filled, tight-binding Peierls-Hubbard
Hamiltonian (PHH) developed \cite{paper_1,paper_2}
to model an interesting class of 1D compounds -- halogen-bridged
transition-metal linear chain complexes ($MX$ chains) --
which exhibit tunable behavior ranging from antiferromagnetic (AF) or
spin-density-wave (SDW) to charge-density-wave (CDW) to semimetallic.
This same model can be considered as: a model of CuO chains
or a 1D analog of the models used for CuO$_2$ planes
in oxide superconductors \cite{MX_CuO};
a 3/4-filled analog of the the organic conductor polyacetylene;
a model for charge-transfer salts such as TTF-TCNQ;
or a model of neutral-ionic transitions \cite{painelli}.
If one considers the two orbitals to be on the same site,
this Hamiltonian is also related to the Kondo Hamiltonian used to
describe heavy fermion and other valence fluctuation materials.
This model yields {\it quantitative} fits for the $MX$ chain compounds
to a variety of experimental data (optical absorption, Raman spectra,
susceptibility data, ESR, $etc.$) \cite{paper_1,paper_2}. Unusual
magnetic behavior in strong magnetic fields has also
been reported in one $MX$ system \cite{haruki_wachter}, which
may be experimental support for the phenomena discussed here.

Our model 1D, 2B, PHH is \cite{paper_1,paper_2}:
\FL
\begin{eqnarray}
H &=&
\sum_{l,\sigma} \biggl\{\bigl(-t_0 +  \alpha \delta_l\bigr)
                \bigl(c^\dagger_{l,\sigma} c_{l+1,\sigma}
                     +c^\dagger_{l+1,\sigma} c_{l,\sigma}\bigr)
\label{E1}\\
&&~~~~~~~~~~~~~~~~~~
+~\bigl[\epsilon_l-\beta_l (\delta_l+\delta_{l-1})\bigr]
               c^{\dagger}_{l,\sigma} c_{l,\sigma} \biggr\}
\nonumber\\
&&+\sum_{l} \biggl\{
U_l n_{l\uparrow} n_{l\downarrow}
+\, {1\over 2} K (\delta_l-a_1)^2
\,+\, P \delta_l
\biggr\} ~,
\nonumber
\end{eqnarray}
where $c^\dagger_{l,\sigma}$ ($c_{l,\sigma}$) creates
(annihilates) an electron at site $l$ with spin $\sigma$,
and $M$ ($d_{z^2}$) and $X$ ($p_z$) Wannier orbitals are situated
on even and odd sites, respectively. Each $M_2X_2$ unit cell has 6
electrons, or 3/4-filling. No non-bonding orbitals were included in
the 1D Hamiltonian, and effective springs were used
to model these and other such elements of the structure not explicitly
included \cite{footnote_params}. Parameters are: the on-site energy
(electron affinity) $\epsilon_l$
($\epsilon_M$=$-$$\epsilon_X$=$e_0$),
electron hopping $(t_0)$, on-site $(\beta_M, \beta_X)$
and inter-site $(\alpha)$ $e$-ph coupling, and
on-site $(U_M, U_X)$ $e$-$e$ repulsion.
$a_1$ is the natural length of the $M$-$X$ spring $K$, and
$\delta_l$ is the relative displacement of the atoms
on an $N$ site chain at sites $l$ and $l$+1.
$\delta_l$ is determined (self-consistently) by minimizing
the total energy $E_T$, $\partial E_T/\partial\delta_l$=0.
The change in average $MX$ bond length $a$, $a$=$\sum_l\delta_l/N$,
and the pressure $P$ are related by $\partial E_T/\partial P$=$Na$,
yielding $P$=$K(a_1$$-$$a$$+$$D)\,$ where
$ D=\sum_{l,\sigma}\bigl\langle 2\beta_l n_{l,\sigma}
 -\alpha (c^\dagger_{l,\sigma} c_{l+1,\sigma}
          +c^\dagger_{l+1,\sigma} c_{l,\sigma})\bigr\rangle/NK $.
We work at fixed $a$.
We have used both mean-field (Hartree-Fock) and exact diagonalization
to study this model \cite{paper_1}. Upon doping, the multi-band
nature is manifested in localized intrinsic defects --
expected on the basis of effective 1B models \cite{baeriswyl_1_band}
to be electron/hole symmetric and chargeless, spinless,
and/or distortionless -- which exhibit
electron/hole asymmetry \cite{gammel_ptcl_90} and local charge, spin,
and/or LD character \cite{paper_1,paper_2}:
for example local LD around ``magnetic" polarons in an AF such as
NiCl or NiBr \cite{paper_2}.  Near phase boundaries, even at low defect
concentration, modification of the background (stoichiometric ``ground
state") can also occur \cite{reichl}.

\begin{floating}[t] \vskip 3.2truecm \narrowtext
\figure{\label{fig1}
$t_0$=0 phase diagram of an infinite chain
for $\beta_X/\beta_M$=$-$1, $U_X/U_M$=1.
The dashed line is the parameter path used
for Fig.~\ref{fig4} and + the point in Fig.~\ref{fig3}.
Note that changing $P$ is equivalent to moving along
a line through the origin.
}
\end{floating}
\begin{floating}[b] \widetext
\widetext
\begin{table}
\caption{
The $a$=$a_1$=0 constant volume period-4 phases, occupancies,
distortions, energies, and effective antiferromagnetic
spin correlations $J$ in the $t_0$$\rightarrow$0 limit.
Here $t$=$t_0$/$e_0$, $u_{m,x}$=$U_{M,X}/e_0$,
$b_{m,x}$=$\beta_{M,X}/\sqrt{Ke_0}$, and
$ \delta(l)\sqrt{K/e_0}$=$ d_X (\cos {l\pi\over2}$$-$$\sin {l\pi\over2} )
          $$-$$d_M (\cos {l\pi\over2}$$+$$\sin {l\pi\over2} ) $
defines dimensionless $X$($M$) sublattice distortion order parameters
$d_X$($d_M$).
}
\smallskip
\begin{tabular}{llccccl}
phase &$X\,M\,X\,M~$
     &$d_M$
     &$d_X$
     &$D\sqrt{K\over e_0}$
     &$2E_T/(e_0N)$
     &$~~~~~~~~~~~~~~~~~$$J^{\rm eff}/e_0$\\
\hline
\smallskip
MAF  &$\!\uparrow\!\!\downarrow$$-$$\uparrow$$-$$\!\uparrow\!\!\downarrow$%
$-$$\downarrow$$-$
     &0
     &0
     &2$b_x$+$b_m$
     &$u_x$$-$1
     &$\hbox{$J_{MM}$}\over\hbox{$e_0$}$=$\hbox{$t^4$(4$u_m$$-$$u_x$+4)}
\over\hbox{$2u_m$($u_m$$-$$u_x$+2)$^2$(2$u_m$$-$$u_x$+4)}$ \\
XAF  &$\uparrow$$-$$\!\uparrow\!\!\downarrow$$-$$\downarrow$$-$%
$\!\uparrow\!\!\downarrow$$-$
     &0
     &0
     &$b_x$+2$b_m$
     &$u_m$+1
     &$\hbox{$J_{XX}$}\over\hbox{$e_0$}$=$\hbox{$t^4$(4$u_x$$-$$u_m$$-$4)}%
\over\hbox{$2u_x$($u_x$$-$$u_m$$-$2)$^2$(2$u_x$$-$$u_m$$-$4)}$ \\
MIX  &$\uparrow$=$\downarrow$$-$$\!\uparrow\!\!\downarrow$$\cdot$$\cdot$%
$\cdot$$\!\uparrow\!\!\downarrow$$-$
     &${1\over2}b_m$
     &${1\over2}b_x$
     &$3{\hbox{$b_x$+$b_m$}\over\hbox{2}}$
     &${\hbox{$u_x$+$u_m$}\over\hbox{2}}$$-$${\hbox{$b_x^2$+$b_m^2$}%
\over\hbox{4}}$
     &$\hbox{$J_{MX}$}\over\hbox{$e_0$}$=$\hbox{$t^2$($u_m$+$u_x$)}%
\over\hbox{($u_x$$-$2+$b_x^2$$-$$b_m^2$)($u_m$+2$-$$b_x^2$+$b_m^2$)}$ \\
CDW  &$\!\uparrow\!\!\downarrow$=$\bullet$=$\!\uparrow\!\!\downarrow$%
$-$$\!\uparrow\!\!\downarrow$$-$
     &$b_m$
     &0
     &2$b_x$+$b_m$
     &$u_x$+${1\over2}u_m$$-$1$-$$b^2_m$
     &$~~~~~~~~~~~~~~~~~~~~$--\\
\smallskip
BOW  &$\bullet$=$\!\uparrow\!\!\downarrow$$-$$\!\uparrow\!\!\downarrow$%
$-$$\!\uparrow\!\!\downarrow$=
     &0
     &$b_x$
     &$b_x$+2$b_m$
     &${1\over2}u_x$+$u_m$+1$-$$b^2_x$
     &$~~~~~~~~~~~~~~~~~~~~$--\\
\end{tabular}
\label{T1}
\end{table}
\narrowtext
\end{floating}

Tuning $e_0$, $t_0$, $U_M$, and $U_X$ is essentially a 1D analog of
theoretical discussions in 2D which focus on the $p$-$d$ hybridization
in HTC materials. There, as here, these parameters determine the broken
symmetry (if any) of the ground state, and the nature of electron
or hole doping into those states. Indeed, the similarity is even
stronger because the nominal filling at stoichiometry
is essentially the same in both $MX$ and oxide HTC materials --
3/4-filling of 2 bands and 5/6-filling of 3 bands, respectively.
Furthermore, both AF and CDW HTC compounds exist
($e.g.$, LaCu$_2$O$_4$, BaBiO$_3$), much as for $MX$ compounds
($e.g.$, NiCl, PtCl, respectively). We stress the $MX$ {\it class}
has the advantage of essentially continuous tunability between these
extremes. Finally, even in the context of AF-based HTC materials the
role of $e$-ph coupling is increasingly appreciated \cite{gen_HTC},
where, upon doping, the same local competition of AF and CDW
as in $MX$ compounds occurs and can induce strong {\it local}
LD in a strongly magnetic background \cite{yonemitsu},
as observed in doping and photodoping spectroscopy \cite{photodoped_HTC}.

\begin{floating}[t] \vskip 2.8truecm \narrowtext
\figure{\label{fig2}
Schematic energy level diagram in the strongly correlated
limit showing frustration of the effective antiferromagnetic couplings
due to the partial real-space occupancy of the low-lying levels
induced by $MX$ hybridization.
}
\end{floating}
\begin{floating}[b] \vskip 6.6truecm \end{floating}

In the CDW case with large LD, typical of many $MX$ chains, a first
approximation is to approach the material as decoupled
$X$=$M$=$X$ trimers and $M$ monomers. Indeed, optical
absorptions in PtCl chains are very close to the monomers and
trimers in solution \cite{PtCl_oligomer}. Similar decoupled-cluster
limits can be formulated for the other LD phases,
though the analysis does not add significantly to the
understanding from the analytic $U$=0 limit \cite{paper_1}
and $t_0$=0 limit (described below).

For large (positive) $e_0$, a reasonable starting
point for interpreting the effects of including $e$-$e$ and $e$-ph
couplings is an effective 1/2-filled, 1B model
focusing on the $M$ $d_{z^2}$ orbitals \cite{baeriswyl_1_band}.
For zero $e$-$e$ correlations, when on-site $e$-ph coupling
($\beta$) dominates, the ground state exhibits $X$-sublattice
LD with an accompanying CDW on the $M$-sublattice
(CDW phase). When the intersite $e$-ph
coupling, $\alpha^{eff}_{1{\rm B}}$$=$$\alpha t_0/(2e_0)$,
dominates, the case is reversed: $X$-CDW and $M$-LD
(bond-order-wave (BOW) phase).
Typical $e$-ph phenomena (phonon softening, solitons, $etc$.)
are found \cite{baeriswyl_1_band}.
The full 2B model leads to qualitative modifications
such as removal of the BOW phase \cite{paper_1}, and electron-hole
asymmetry of defect absorptions \cite{gammel_ptcl_90}.
Our focus here is on intermediate to large $e$-$e$ correlations
(large $U_M$), where in the 1D, 1B limit one expects a SP phase with an
effective AF coupling between $M$ sites \cite{zhang_rice,emery,shi_t_J}.

\begin{floating}[t]  \vskip 3.2truecm \narrowtext
\figure{\label{fig3}
$t_0$ dependence of the difference in singlet and triplet
energies for the CDW, MIX, and MAF phases
at the point indicated by the + in Fig.~\ref{fig1},
obtained by exact diagonalization of an $N$=16 site system.
For small $t_0$, this corresponds to excitations
of the effective spin-Hamiltonian $H=4JS^z_i S^z_{i+j}$
with energy $4J_{MM}$ in the MAF phase and $2J_{MX}$ in
the MIX phase, whereas in the CDW phase this reflects
the Peierls gap. The dotted lines are the perturbative
expressions for $J_{MM}$,$J_{MX}$ from Table \ref{T1}.
}
\end{floating}

To understand the consequences of $e$-$e$ correlations
for a multiband model, we first examine the zero-hopping limit. The
period-4 (P4) phases for $t_0$=$\alpha$=0 are listed in Table \ref{T1}.
The phase diagram for parameter values
near the CDW/AF crossover (relevant to $M$I or Ni$X$ materials),
is shown in Fig.~\ref{fig1}. The phase diagram is more complex for
$U_X$,$|\beta_X|$$\agt$$U_M$,$|\beta_M|$ and/or $\beta_X$/$\beta_M$$>$0,
where the hybridization-driven competition is most effective.
When $t_0$$\equiv$0, all spin excitations are isoenergetic.
For $t_0$$\ne$0, AF correlations develop and one can treat the
problem in terms of an effective spin model.
Without coupling between the two bands, the lower, $X$-like
(for $e_0$$>$0) band is full and non-magnetic, while the
upper, $M$-like band is 1/2-full with one electron per $M$
site (when $e$-$e$ repulsion is dominant). The on-site $e$-ph
coupling $\beta$ leads to a splitting of the upper band which
competes with an AF ordering of the spins caused by the
effective AF coupling between neighboring metal sites, $J_{MM}$
(as in the effective 1B case).
When hybridization between the two bands is allowed,
the lower band is not completely full, and now there is
an effective AF coupling between neighboring halide sites,
$J_{XX}$ (dominant when $U_X$ is dominant),
as well as $M$ and $X$ sites, $J_{MX}$. $J_{XX}$ and $J_{MX}$
are not present in 1B models \cite{1bcaveat}. In fact,
when the splitting due to the $e$-ph coupling $\beta$
is on the order of $U$ and $e_0$, the AF state with neighboring
$M$-$X$ pairs singly occupied can become the ground state, as shown
in Fig.~\ref{fig1}. This implies, in contrast to the 1B case, that
the combination of $e$-$e$ and $e$-ph coupling in the 2B model drives,
in addition to the non-magnetic CDW and BOW phases, three (competing)
SP phases: one on the $X$-sublattice (XAF), one on the $M$-sublattice
(MAF), and one involving $MX$ pairs and large LD (MIX).
Since $J_{MM}$, $J_{XX}$, and $J_{MX}$ are all AF couplings
($J>0$), they obviously cannot all be simultaneously satisfied
and the system is frustrated,  as shown in Fig.~\ref{fig2}.
It is straightforward to derive from fourth-order perturbation
theory in $t_0$ an effective $t$, $J_{MX}$, $J_{MM}$, $J_{XX}$ model
in the large $U$ limit; the resultant $J$'s are listed in Table \ref{T1}.
When only one of the $J$'s dominates, one can numerically check this
estimate by comparing the energies of the singlet and triplet ground states
(at fixed LD).  Fig.~\ref{fig3} shows good agreement for small $t_0$.
Note that the CDW phase has an entirely $e$-ph driven AF component:
even when the $X(M)$-LD
is large, some residual $M(X)$-sublattice magnetization remains.

\begin{floating}[t] \vskip 7truecm \narrowtext
\figure{\label{fig4} Exact diagonalization:
(a) lattice distortion amplitude, (b) total energy $E_T$, and
(c) n.n. and (d) n.n.n. spin correlations
as a function of $U_M/e_0$ for $t_0/e_0$=0.5
and N=16 for MAF, CDW, and MIX phases.
Other parameters are as in Fig.~\ref{fig1}.
The charge densities have essentially the $t_0$=0
values shown schematically in Table \ref{T1}.
}
\end{floating}

For parameters where the LD is large and driven by the
on-site $e$-ph coupling $\beta$, and/or the Hubbard $U$ terms
are large, the zero-hopping phase diagram is close
to the actual phase diagram.  In Fig.~\ref{fig4}
we show the LD, total energy, and spin correlations
obtained by exact diagonalization on a ring of 16 sites
as parameters are varied along the line shown in Fig.~\ref{fig1}.
As $U_M$ increases (LD decreases), we observe a sharp transition from
$M$-$M$ charge coupling (CDW) to $M$-$X$ spin coupling (MIX)
to $M$-$M$ spin coupling (MAF), in agreement with the $t_0$=0
results. (Since for finite systems the phases are strongly pinned
by the LD, we can follow competing phases across phase boundaries.)
For larger $t_0$, the couplings change more smoothly,
and charge disproportionation, $M$-$X$ spin coupling,
and $M$-$M$ spin coupling coexist, with the phase of the LD
passing from CDW through MIX to BOW as the amplitude goes to zero.
For large $t_0$, even for $U_M$=0, the CDW phase shows
a strong tendency towards AF order,
driven by valence fluctuations rather than $U$.

The crossover from P4 CDW to P4 MAF is
accompanied by LP SL phases \cite{batistic_diff_than_J}
when $|\beta_X/\beta_M|$$>$1, or when the
{\it intersite} $e$-ph coupling $\alpha$ is large.
We have completed only preliminary evaluation
of the LP region of the phase diagram, as the many competing
metastable states significantly complicate the analysis.
An $\alpha$-driven SL phase has recently
been found in a 2D, 3B model of HTC CuO$_2$ layers \cite{yonemitsu}.
Such SLs may be viewed as an ordered array of
discommensuration defects with respect to a nearby commensurate
order ($e.g.$, P4). In view of the effective $J$'s discussed
above, it may be natural to model such states in terms of ANNNI-like
models \cite{annni,emery,imada}, where nearest and longer range couplings
compete, leading to frustration and associated complexity
phenomena -- multitimescale relaxation, hysteresis, metastability,
$etc$. In the context of the $MX$ class, it will be
particularly interesting to investigate materials in, or near,
this crossover regime -- $e.g.$, PtI -- and to further control
the crossover with pressure, magnetic field,
doping, impurities, $etc$. Indeed {\it doping} into this
complex regime should be highly sensitive to the softness
and competitions of the phases: this is an excellent
regime to study pairing tendencies and metallization.

In conclusion, we stress that the 1D, 2B, PHH is representative
of a very large variety of low-D electronic materials. This variety
is mirrored in the model's richness, not only in terms of the
potential ground states as discussed here, but also in terms of the
consequences of doping into such phases, especially near phase boundaries,
where besides the usual plethora of doping and photoinduced
non-linear excitations (solitons, polarons, bipolarons, and
excitons \cite{paper_1,paper_2,painelli,baeriswyl_1_band}),
a dopant-induced transition
of the $global$ phase may exist \cite{reichl,photodoped_HTC}
and novel pairing mechanisms are
anticipated \cite{zhang_rice,emery,yonemitsu,imada}.
Inclusion of $e$-ph in a 2D 3B model relevant to HTC materials
leads to similar effects both at stoichiometry and upon doping,
where generalized polaronic or ``bag" states with local coexistence
of spin and charge are found \cite{yonemitsu}.
We are attempting to exploit this richness via a systematic
``making-measuring-modeling" approach on to the $MX$ class of 1D
compounds, which are realizations of this Hamiltonian spanning the
range of complex broken symmetry ground states. The recent device
proposals for photovoltaics, $etc.$, \cite{saxena_devices} serve
to underscore the exciting consequences of this richness.
We believe that experimental investigations of the
template \cite{template} and pressure dependence and high (magnetic) field
behavior of pure and doped $MX$ materials in the LD/AF cross-over regime,
such as PtI, NiBr, or their mixed-metal or -halide analogs,
will continue to yield interesting insights into the nature
of intrinsically multiband effects and the competition
between $e$-$e$ and $e$-ph interactions.


{\it Acknowledgements}
We thank with X.Z.~Huang, A.~Saxena, J.~Shi, and K.~Yonemitsu,
for important discussions, and E.Y.~Loh, Jr.~for computational assistance.
ARB was supported by the US DOE,
HR by the Deutsche Forschungsgemeinschaft through SFB 213,
and JTG by a National Research Council-NRaD Research Associateship
through a grant from the ONR.
Supercomputer access was through the Advanced Computing
Laboratory at LANL. HR and JTG are grateful
to the hospitality of LANL, where this work was begun.



\end{narrowtext}

\begin{references}
\bibitem{footnote_params}
Longer-range Coulomb effects have been investigated \cite{painelli}.
The $e$-ph couplings $\alpha$ and $\beta$ may
be viewed as expansions of longer-range Fr\"ohlich-like terms.
We have also tested further neighbor hoppings and springs.
Such extensions do not change the qualitative discussion here.
\bibitem{1bcaveat}
While effective 1B $t$-$J$ models may be valid in both the MAF and MIX
phases -- $e.g.$ using $MX$ singlets \cite{zhang_rice} potentially
with an $e$-ph dependent $J$ \cite{imada} --
to model the MAF/MIX crossover would require
distinguishing between $MX$=2$\uparrow$ and $\uparrow$2,
and thus a multiorbital model, as at the crossover one can
no longer argue there is a separation of energy scales.
\bibitem{batistic_diff_than_J}
Intermediate phases can be driven by the competitions
between $t_0$, $\beta$, $\alpha$, long-range Coulomb fields,
and/or the effective short range $J$'s.
Such intermediate phases
can be P4, as expected from commensurability effects, or LP, since
frustration can drive effective long-range interactions,
as seen in ANNNI models \cite{annni}.
The large-$\beta$ LP phase
discussed in Ref.~ \cite{batistic_long_per_91}, though related,
is somewhat different than the frustration driven, magnetic LP phase
discussed here or, in 2D, in Ref.~\cite{yonemitsu} (for large $\alpha$).
\bibitem{template} In mixed $MXX^\prime$ crystals, the $MX^\prime$
phase may be controlled
by the lattice constant of the host $MX$ crystal.
\end{references}
\end{document}